\documentclass[12pt]{article}
\usepackage{amsmath,amsfonts}
\usepackage{amssymb}
\usepackage{amsthm}
\usepackage{times,txfonts}

\def\largesymbol#1{\mbox{\strut\rlap{\smash{\Large$#1$}}\quad}}

\newtheorem{thm}{Theorem}[section]
\newtheorem{prop}[thm]{Proposition}

\newtheorem{remark}[thm]{\it Remark}
\newtheorem{example}[thm]{\it Example}

\numberwithin{equation}{section}

\def\pf{\noindent{\it Proof.} \ }

\def\qed{\hfill $\square$}

\title{On an integrable system of $q$-difference equations satisfied by the universal characters: its Lax formalism and an application to $q$-Painlev\'e equations}
\author{Teruhisa TSUDA   \\
Faculty of Mathematics, Kyushu University, \\  
Hakozaki, Fukuoka 812-8581, Japan.  \\
tudateru@math.kyushu-u.ac.jp}
\date{}

\begin{document}
\maketitle

\begin{abstract}
The universal character is a generalization of the Schur function attached to a pair of partitions.
We study an integrable system of $q$-difference equations satisfied by the universal characters,
which is 
an extension of the $q$-KP hierarchy and is
called the {\it lattice $q$-UC hierarchy}.
We describe the lattice $q$-UC hierarchy as a compatibility condition of its
associated linear system (Lax formalism) and 
explore an application to the $q$-Painlev\'e equations via similarity reduction.
In particular a higher-order analogue of the $q$-Painlev\'e VI equation is presented.
\end{abstract}

\section{Introduction}

The universal character, defined by Koike \cite{koi},
is a polynomial attached to a pair of partitions (Young diagrams) and is
a generalization of the Schur function.
The universal character describes the  irreducible rational character 
of the general linear group,
while the Schur function, as is well known, does the irreducible polynomial character of the group.
The Schur function is significant also for
the study of integrable systems;
in fact, Sato \cite{sat} showed that the KP hierarchy, an important class of  
nonlinear partial differential equations of soliton type,
is exactly an infinite-dimensional integrable system characterized by the Schur function.
Needless to say,
the KP hierarchy provides a basic prototype
in the field of integrable systems
involving the Painlev\'e equations.
Furthermore, 
variants of  the KP hierarchy
including
difference and $q$-difference versions 
have been extensively studied as well as the original one.

On the other hand, 
in  \cite{t04} the author proposed an extension of the KP hierarchy, 
called the {\it UC hierarchy},
as an infinite-dimensional integrable system characterized by the universal character.
For instance, the whole set of homogeneous polynomial solutions of the UC hierarchy 
is in one-to-one correspondence with the set of the universal characters.
Also in \cite{t05b}
a $q$-difference analogue of the hierarchy 
was studied in connection with the $q$-Painlev\'e equations;
it is an integrable system of $q$-difference equations
originated from certain quadratic relations among the universal characters
and was named the {\it lattice $q$-UC hierarchy} ($q$-LUC),
since  whose
dependent variables ($\tau$-functions) are arranged on a two-dimensional lattice ${\mathbb Z}^2$; {\it cf}. \cite{t05a,tm}.
However, some basic properties, {\it e.g.}, Lax formalism, 
of $q$-LUC as an integrable system have been unclear so far.

The aim of this paper is to provide an alternative formulation of 
$q$-LUC and explore its application to the $q$-Painlev\'e equations.
In Sect.~\ref{sect:lax},
we first reformulate $q$-LUC as a compatibility condition of 
its auxiliary system of linear equations (Lax formalism).
Based on the Lax formalism, 
we then describe $q$-LUC as a system of $q$-difference evolution equations
for appropriate dependent variables,
which naturally includes the $q$-KP hierarchy (see, {\it e.g.}, \cite{kny02b}) as a special case.
Next, in Sect.~\ref{sect:q-p}  we consider 
a reduction of $q$-LUC by requiring its solutions to satisfy  
certain homogeneity and periodicity
(a similarity reduction). 
As a result, we obtain a class of 
invertible, 
or rather, {\it birational} 
discrete dynamical systems 
of $q$-Painlev\'e type.
For example, we present 
a higher-order extension of the $q$-Painlev\'e VI equation
in Sect.~\ref{subsect:j=0}. 
Observing the universal characters be consistent with the similarity reduction,
we can immediately construct
particular solutions of the dynamics in terms of the universal characters;
{\it cf}. \cite{t05b}.

\section{Universal characters and $q$-difference integrable systems}
\label{sect:lax}

We begin with recalling the definition of the universal characters.
The lattice $q$-UC hierarchy ($q$-LUC)
takes the form of bilinear $q$-difference equations 
that arises from quadratic relations among the universal characters.
In this section we present its associated Lax formalism; 
that is, we reformulate $q$-LUC
as a compatibility condition of an auxiliary system of linear 
$q$-difference equations.

\subsection{Universal character and lattice $q$-UC hierarchy}

A {\it partition} $\lambda=(\lambda_1, \lambda_2, \ldots )$
is a sequence of non-negative integers such that
$\lambda_1 \geq \lambda_2 \geq \cdots \geq 0$
and that $\lambda_i =0$ for $i \gg 0$.  
The number of $\{ i \, | \, \lambda_i \neq 0\}$ is called the {\it length} of $\lambda$.
For a pair of partitions $\lambda, \mu$,
the {\it universal character} $S_{[\lambda,\mu]}$ is 
a polynomial in (infinitely many) variables 
${\boldsymbol x}=(x_1,x_2,\ldots)$ and ${\boldsymbol y}= (y_1,y_2, \ldots)$
and is defined by the {\it twisted} Jacobi-Trudi formula:
\begin{equation}  \label{eq:def-of-uc}
S_{[\lambda,\mu ]}({\boldsymbol x},{\boldsymbol y})
= \det 
\left(
  \begin{array}{ll}
 p_{\mu_{l'-i+1}  +i - j }({\boldsymbol y}),  &  1 \leq i \leq l'  \\
 p_{\lambda_{i-l'}-i+j}({\boldsymbol x}),     &  l'+1 \leq i \leq l+l'   \\
  \end{array}
\right)_{1 \leq i,j \leq l+l'},
\end{equation}
with $l=l(\lambda)$ and $l'=l(\mu)$.
Here   
$p_n({\boldsymbol x})$ $(n \in {\mathbb Z})$
is determined by 
the generating function
$\sum_{n=0}^\infty p_n({\boldsymbol x})z^n =  
\exp 
\left(  \sum_{n=1}^\infty x_n z^n  \right)$
and $p_n=0$ for $n <0$.
In the case where $\mu=\emptyset$, we have
\[S_{[\lambda, \emptyset ]}({\boldsymbol x},{\boldsymbol y})=\det \bigl( p_{\lambda_i-i+j}({\boldsymbol x}) \bigr) =: S_\lambda({\boldsymbol x}),
\]
which is exactly the Schur function.
If we count the degree of each variable as
$\deg x_n=n$ and
$\deg y_n=-n$,
then 
$S_{[\lambda, \mu ]}({\boldsymbol x},{\boldsymbol y})$ 
is a homogeneous polynomial of degree 
$|\lambda| - |\mu|$,
where $|\lambda|=\lambda_1+\cdots+\lambda_l$.
The universal character 
$S_{[\lambda,\mu ]}$
describes the irreducible character of a rational representation of 
the general linear group
corresponding to a pair of partitions $[\lambda,\mu]$,
while the Schur function 
$S_\lambda$
does
that of a polynomial representation
corresponding to a partition $\lambda$;
see \cite{koi} for details.

\begin{example}\rm
We have $p_0=1$, $p_1({\boldsymbol x})=x_1$,
$p_2({\boldsymbol x})=x_2+x_1^2/2$,
$p_3({\boldsymbol x})=x_3+x_2x_1+x_1^3/6$, and so on.
When $\lambda=(2,1)$ and $\mu=(1)$,
the universal character reads
\[
S_{[(2,1),(1)]}({\boldsymbol x},{\boldsymbol y})= 
\left|  
 \begin{array}{ccc} 
  p_1({\boldsymbol y}) & p_0({\boldsymbol y})& p_{-1}({\boldsymbol y})  \\
  \hline
  p_1({\boldsymbol x}) & p_2({\boldsymbol x})& p_3({\boldsymbol x})  \\
  p_{-1}({\boldsymbol x}) & p_0({\boldsymbol x})& p_1({\boldsymbol x})  
 \end{array} 
\right|
=\left( \frac{x_1^3}{3}-x_3\right)y_1-x_1^2,
\]
which is a homogeneous polynomial of degree $|\lambda|-|\mu|=3-1=2$.
\end{example}

Let $I\subset {\mathbb Z}_{>0}$ and $J\subset {\mathbb Z}_{<0}$
be finite indexing sets and
$t_i$ $(i \in I \cup J)$ the independent variables.
Let $T_i=T_{i;q}$ be the $q$-shift operator defined by
\[
T_{i;q}(t_i) = \left\{ 
\begin{array}{ll}
q t_i & (i \in I) \\
q^{-1} t_i &(i \in J)
\end{array} 
\right.
\quad
\text{and}
\quad
T_{i;q}(t_j)=t_j  \quad (i \neq j).
\]
For unknown functions
$\tau_{m,n}=\tau_{m,n}({\boldsymbol t})$ in $t_i$ 
arranged on the lattice 
$(m,n) \in {\mathbb Z}^2$,
the following system of $q$-difference equations
is called the {\it lattice $q$-UC hierarchy} ($q$-LUC):
\begin{equation}  \label{eq:bil}
t_i
T_{ i} (\tau_{m,n+1})
T_{ j } (\tau_{m+1,n}) 
-t_j
T_{ j }  (\tau_{m,n+1})
T_{ i } (\tau_{m+1,n}) 
=(t_i-t_j)
T_{i}T_{j}(\tau_{m,n})
\tau_{m+1,n+1} 
\end{equation}
where
$i,j \in I \cup J$.

The $q$-LUC was introduced in \cite{t05b} and
it is, as seen below, 
originated from the quadratic relations satisfied by the universal characters.
For convenience,
we shall extend the universal character,
(\ref{eq:def-of-uc}),
to be defined for a pair of arbitrary sequences of integers $[\lambda,\mu]$.
Note that one can associate a unique partition $\hat\lambda$
with any given sequence of integers $\lambda$ 
by applying  successively
a  procedure like 
$(\ldots, k, l, \ldots) \mapsto (\ldots, l-1,k+1, \ldots)$;
we have 
$S_{[\lambda,\mu]}=\pm S_{[\hat\lambda,\mu]}=\pm S_{[\lambda,\hat\mu]}$.
Define the function 
$s_{[\lambda,\mu]}=s_{[\lambda,\mu]}({\boldsymbol t})$  by
\[
s_{[\lambda,\mu]}({\boldsymbol t})=S_{[\lambda,\mu]}({\boldsymbol x},{\boldsymbol y})
\]
via
the change of variables
\begin{equation}  \label{eq:cha}
x_n = \frac{\sum_{i \in I} t_i^n - q^n \sum_{j \in J}t_j^n }{n(1-q^n)}
\quad \text{and} \quad
y_n = \frac{\sum_{i \in I} t_i^{-n} - q^{-n} \sum_{j \in J}t_j^{-n} }{n(1-q^{-n})}.
\end{equation}

\begin{prop}[\cite{t05b}]  \label{prop:uc}
We have
\[
t_i
T_{ i} (s_{[\lambda,(k',\mu)]} )
T_{ j } (s_{[(k,\lambda),\mu]}) 
-t_j
T_{ j } (s_{[\lambda,(k',\mu)]} )
T_{ i } (s_{[(k,\lambda),\mu]} )
=(t_i-t_j)
T_{i}T_{j}(s_{[\lambda,\mu]})
s_{[(k,\lambda),(k',\mu)]}
\]
for any integers $k,k'$ and partitions $\lambda, \mu$.
\end{prop}

\begin{remark}[Symmetry of $q$-LUC]\rm  
\label{rem:sym}
If a set of functions
$\{\tau_{m,n}({\boldsymbol t})\}_{m,n}$ is a solution of $q$-LUC, (\ref{eq:bil}),
so is $\{ c^{-d_{m,n}}\tau_{m,n}(c {\boldsymbol t})\}_{m,n}$ for arbitrary constants
$c \in {\mathbb C}^*$ and $d_{m,n} \in {\mathbb C}$ 
with
$d_{m,n}+d_{m+1,n+1}=d_{m,n+1}+d_{m+1,n}$.
Accordingly it seems reasonable to take a particular interest in the fixed solutions with respect to such a scaling symmetry.
For example, 
the universal character $s_{[\lambda,\mu]}$ gives a homogeneous solution of $q$-LUC
as seen from Prop.~\ref{prop:uc},
and by definition it  satisfies
$s_{[\lambda,\mu]}(c {\boldsymbol t})=c^{|\lambda|-|\mu|}  s_{[\lambda,\mu]}({\boldsymbol t})$.
Such self-similar solutions will be investigated in Sect.~\ref{sect:q-p} below.
\end{remark}

\begin{remark} 
\rm  
Suppose $|I|+|J| \geq 3$. 
Then it can be verified from (\ref{eq:bil}) 
as a {\it necessary} condition
that
\begin{align}  
&(t_i-t_j)T_{i}T_j(\tau_{m,n})T_k(\tau_{m+1,n})
+(t_j-t_k)T_jT_k(\tau_{m,n})T_i(\tau_{m+1,n})
\nonumber
\\
&+(t_k-t_i)T_{i}T_k(\tau_{m,n})
T_j(\tau_{m+1,n})
=0
\label{eq:q-uc-old}
\end{align}
for any $i,j,k \in I \cup J$.
This, (\ref{eq:q-uc-old}),
is nothing but the bilinear equation introduced previously in \cite{t05a} 
and named the {\it $q$-UC hierarchy} (though this terminology is a little confusing).
\end{remark}

\subsection{Lax formalism}
In order to derive the associated linear system from the bilinear equation (\ref{eq:bil}) of $q$-LUC,
we need to prepare an extra variable $z$ called the {\it spectral parameter}.
Through the change of variables (\ref{eq:cha}),
we regard $\tau_{m,n}({\boldsymbol t})$ as a function in $({\boldsymbol x},{\boldsymbol y})$ 
and write 
$\tau_{m,n}({\boldsymbol t})=\widetilde{\tau}_{m,n}({\boldsymbol x},{\boldsymbol y})$.
Let us now introduce the 
{\it wave function} $\psi_{m,n}=\psi_{m,n}({\boldsymbol t},z)$,
a function in $z$ and $t_i$ 
$(i \in I \cup J)$,
defined by
\begin{equation}
\psi_{m,n}
= \frac{1}{(-z)^m \prod_{i \in I} (z^{-1} t_i; q)_\infty  \prod_{j \in J} (z^{-1}t_j; q^{-1})_\infty} 
\frac{\widetilde{\tau}_{m,n}({\boldsymbol x}-[z],{\boldsymbol y}-[z^{-1}])}{\widetilde{\tau}_{m,n+1}({\boldsymbol x},{\boldsymbol y}) },
\end{equation}
where we adopt the following convention of
{\it $q$-shifted factorial}:
\[
(a;q)_\infty=\prod_{i=0}^{\infty}(1-a q^i),
\]
and $[z]=(z,z^2/2,z^3/3,\ldots)$.
In view of (\ref{eq:bil}),
we see that
the wave function solves the linear system of $q$-difference equations:
\begin{equation} \label{eq:lax}
T_{i}(\psi_{m,n})=u_{i,m,n} \psi_{m,n+1}+t_i \psi_{m+1,n}
\quad (i \in I \cup J)
\end{equation}
where
\begin{equation} \label{eq:def-u}
u_{i,m,n}= \frac{ T_i(\tau_{m+1,n}) \tau_{m,n+2}}{ T_i(\tau_{m,n+1}) \tau_
{m+1,n+1}}.
\end{equation}

\begin{prop}  \label{prop:u}
The compatibility condition $T_i T_j=T_jT_i$ 
of (\ref{eq:lax})
yields 
\begin{equation} \label{eq:u}
T_j(u_{i,m,n})=u_{i,m,n+1} 
\frac{t_i u_{j,m+1,n} - t_j u_{i,m+1,n}}{t_i u_{j,m,n+1} - t_j u_{i,m,n+1}}
\quad (i \neq j),
\end{equation}
the nonlinear evolution equations for variables 
$u_{i,m,n}=u_{i,m,n}({\boldsymbol t})$.
\end{prop}

\pf
We will write the $q$-shift of a function $F=F({\boldsymbol t})$ 
as
$T_{i_1}T_{i_2} \cdots T_{i_r} (F)=:F^{(i_1,i_2,\ldots,i_r)}$ for brevity.
Applying $T_j$ to (\ref{eq:lax}),
we have
\begin{align*}
T_jT_i(\psi_{m,n})
&=u_{i,m,n}^{(j)} \psi_{m,n+1}^{(j)}+t_i \psi_{m+1,n}^{(j)}
\\
&=u_{i,m,n}^{(j)} ( u_{j,m,n+1} \psi_{m,n+2} +t_j \psi_{m+1,n+1})
+t_i ( u_{j,m+1,n} \psi_{m+1,n+1} +t_j \psi_{m+2,n})
\\
&= u_{i,m,n}^{(j)} u_{j,m,n+1} \psi_{m,n+2}
+ (t_j u_{i,m,n}^{(j)} + t_i u_{j,m+1,n}  ) \psi_{m+1,n+1} +t_it_j \psi_{m+2,n}.
\end{align*}
Hence the compatibility condition
 $T_i T_j=T_jT_i$ 
 is equivalent to 
 \begin{subequations} \label{subeq:comp}
\begin{align}
u_{i,m,n}^{(j)} u_{j,m,n+1} &= u_{j,m,n}^{(i)} u_{i,m,n+1},
\label{eq:comp1} \\
t_j u_{i,m,n}^{(j)}+t_i u_{j,m+1,n} &=t_i u_{j,m,n}^{(i)}+t_j u_{i,m+1,n}.
\label{eq:comp2}
\end{align}
\end{subequations}
Accordingly (\ref{eq:u}) holds.
\qed
\\

Conversely, we shall deduce the bilinear form (\ref{eq:bil}) of $q$-LUC
from the compatibility condition
(\ref{subeq:comp})
of the linear system (\ref{eq:lax}).
Taking an auxiliary variable
\begin{equation} \label{eq:def-w}
w_{m,n}= \frac{\tau_{m+1,n}}{\tau_{m,n+1}},
\end{equation}
we can then write (see (\ref{eq:def-u}))
\[u_{i,m,n}= \frac{T_i(w_{m,n})}{w_{m,n+1}}.
\]
At the level of variables $w_{m,n}$ 
the first condition (\ref{eq:comp1}) reduces to trivial
and the second
(\ref{eq:comp2}) becomes
\begin{equation}  \label{eq:w}
w_{m,n}^{(i,j)} = \frac{w_{m,n+1}^{(i)}w_{m,n+1}^{(j)}}{ w_{m+1,n+1} }
\frac{t_i w_{m+1,n}^{(j)} - t_j w_{m+1,n}^{(i)} }{t_i w_{m,n+1}^{(j)} - t_j w_{m,n+1}^{(i)}}.
\end{equation}
Substituting (\ref{eq:def-w}) into (\ref{eq:w}), we obtain 
\[
\frac{  t_i \tau_{m+1,n+1}^{(i)} \tau_{m+2,n}^{(j)}-t_j \tau_{m+1,n+1}^{(j)} \tau_{m+2,n}^{(i)} }{\tau_{m+1,n}^{(i,j)}  \tau_{m+2,n+1}}
=\left. \text{(LHS)} \right|_{(m,n) \mapsto (m-1,n+1)}.
\]
Consequently the above formula can be decomposed as
\[
 t_i \tau_{m+1,n+1}^{(i)} \tau_{m+2,n}^{(j)}-t_j \tau_{m+1,n+1}^{(j)} \tau_{m+2,n}^{(i)}
 = \alpha(m,n)\tau_{m+1,n}^{(i,j)}  \tau_{m+2,n+1},
\]
where $\alpha(m,n)$ being an arbitrary function such that $\alpha(m-1,n+1)=\alpha(m,n)$.
Assume $\tau_{m,n} \equiv 1$ (for $\forall m,n$) to be a solution. 
Then we get $\alpha(m,n)=t_i-t_j$
and arrive at (\ref{eq:bil}) as desired.

\begin{remark}[Comparison with $q$-KP hierarchy]
\label{rem:kp}
\rm

As predicted from the fact that the universal character is a generalization of
the Schur function,
the lattice $q$-UC hierarchy 
gives a natural extension of the $q$-KP hierarchy.
Consider the case where
$\tau_{m,n}({\boldsymbol t})$ does not depend on $n$;
thus, $\psi_{m,n}$ and $u_{i,m,n}$ do not also.
Rename the dependent variables as 
\begin{equation}  \label{eq:var-kp}
\rho_m := \tau_{m,n}, \quad \phi_m := \psi_{m,n} \quad
\text{and} \quad 
v_{i,m} := \frac{T_i(\rho_{m+1}) \rho_m}{ T_i(\rho_m) \rho_{m+1}} 
=u_{i,m,n} 
\end{equation}
for avoiding confusion.
Then (\ref{eq:bil}), (\ref{eq:lax}) and (\ref{eq:u}) 
reduce respectively to 
\begin{equation}
t_i
T_{ i} (\rho_{m})
T_{ j } (\rho_{m+1}) 
-t_j
T_{ j }  (\rho_{m})
T_{ i } (\rho_{m+1}) 
=(t_i-t_j)
T_{i}T_{j}(\rho_{m})
\rho_{m+1},
\end{equation}
\begin{equation}  \label{eq:kp-lax}
T_{i}(\phi_{m})
=v_{i,m} \phi_{m}+t_i \phi_{m+1},
\end{equation}
and
\begin{equation}
T_j(v_{i,m})
=v_{i,m} 
\frac{t_i v_{j,m+1} - t_j v_{i,m+1}}{t_i v_{j,m} - t_j v_{i,m}},
\end{equation}
 the bilinear form, 
 the associated linear system, 
 and the nonlinear expression
 of the $q$-KP hierarchy; {\it cf.} \cite{kny02b}.
 \end{remark}

\begin{remark}[Lax matrices of $q$-LUC with $(M,N)$-periodicity]
\rm
Suppose that the $(M,N)$-periodic condition: 
$\tau_{m,n}=\tau_{m+M,n}=\tau_{m,n+N}$
holds.
Take
an $MN$-vector 
\[
\Psi={}^{\rm T}\left(\psi_{1,1}, \psi_{2,1}, \ldots, \psi_{M,1},\psi_{1,2}, \psi_{2,2}, \ldots, \psi_{M,2}, \ldots , \psi_{1,N}, \psi_{2,N}, \ldots, \psi_{M,N}
\right).
\]
The linear equation (\ref{eq:lax}) can be therefore
rewritten into the matrix equation:
\begin{equation}  \label{eq:matlax}
T_i (\Psi)= B_i \Psi,
\end{equation}
where
\[
B_i= 
\left(
\begin{array}{ccc|cccc}
&& & u_{i,1,1} & & & \\
\multicolumn{3}{c|}{ \largesymbol{O}_{ M(N-1)\times M }} && u_{i,2,1}& & \\
&&&&& \ddots & \\
&&&&&&   u_{i,M,N-1}\\
\hline
u_{i,1,N} &&&&&&\\
& \ddots &&
\multicolumn{4}{|c}{ \largesymbol{O}_{M \times  M(N-1)} } \\
&&  u_{i,M,N} &&&&
\end{array}
\right)
+t_i 
\left(
\begin{array}{ccc}
\Lambda &&\\
&\ddots&\\
&&\Lambda
\end{array}
\right)
\]
and 
\[
\Lambda=
\left. 
\left(
\begin{array}{cccc}
0& 1&&\\
&\ddots& \ddots &\\
&& &1 \\
(-z)^{-M}
&&&0
\end{array}
\right) \right\} \text{\footnotesize $M$}
.
\]
The compatibility condition of (\ref{eq:matlax}) is expressed as 
\begin{equation}
T_i(B_j)B_i=T_j(B_i)B_j.
\end{equation}
If $N=1$,  the situation above reduces to that of the $q$-KP hierarchy again.
\end{remark}

\section{Associated birational dynamics and $q$-Painlev\'e equations}
\label{sect:q-p}

While the equation (\ref{eq:u}) given in Prop.~\ref{prop:u}: 
\[
T_j(u_{i,m,n})=u_{i,m,n+1} 
\frac{t_i u_{j,m+1,n} - t_j u_{i,m+1,n}}{t_i u_{j,m,n+1} - t_j u_{i,m,n+1}}
\quad (i \neq j),
\]
describes a time evolution ($q$-shift) of the variable $u_{i,m,n}$ 
with respect to $t_j$, 
there are some problems: 
\begin{center}
\begin{tabular}{l}
(i) the time evolution for $i = j $ is undefined; 
\\
(ii) the inverse transformation $T_i^{-1}$ is also undefined.
\end{tabular}
\end{center}

In this section we demonstrate how to settle these problems
by means of the view point of 
similarity reduction. 
To be accurate, 
if we impose some homogeneity and periodicity on the dependent variables (appropriately chosen), we can describe 
the time evolution in terms of invertible or rather {\it birational} mappings.
Interestingly enough, the resulting discrete dynamical systems give rise to 
$q$-difference Painlev\'e equations.

\subsection{The case of $q$-KP hierarchy with $J=\emptyset$}
\label{subsect:qkp}

As a prototypical example,
we first review the result \cite{kny02b}
in the case of the $q$-KP hierarchy 
($q$-KP).
We will continue on the convention used in Remark~\ref{rem:kp}.

Let $I=\{1,2, \ldots, L \}$ and $J= \emptyset$.
Impose on the variables
$\rho_m=\rho_m({\boldsymbol t})$ 
the $M$-periodic condition: 
$\rho_{m+M}=\rho_{m}$
and the homogeneity condition:
$\rho_{m}(q {\boldsymbol t})=q^{d_m}  \rho_m({\boldsymbol t})$
($d_m \in {\mathbb C}$).
Concerning the variables $v_{i,m}$,
the constraint above implies that
\[
v_{i,m+M}=v_{i,m}
\quad \text{and} \quad
T_{1}T_{2} \cdots T_{L}(v_{i,m})=v_{i,m}. 
\]
In parallel,
the wave function 
$\phi_m=\phi_m({\boldsymbol t},z)$
comes to satisfy
$(-z)^{M}\phi_{m+M}=\phi_m$
and
$q^{m} \phi_m(q {\boldsymbol t},q z)=\phi_m({\boldsymbol t},z)$.
Now take the dependent variables 
\begin{equation}
x_{i,m}=x_{i,m}({\boldsymbol t}):= t_i^{-1} T_{i+1}T_{i+2} \cdots T_{L}(v_{i,m})
\quad
(i \in I, m \in {\mathbb Z}/ M {\mathbb Z})
\end{equation}
and
let ${\mathbb C}({\boldsymbol x})$ be the field of rational functions in variables $x_{i,m}$.

\begin{thm}[Kajiwara--Noumi--Yamada \cite{kny02b}] \label{thm:kny}
The action of $T_j$ on variables $x_{i,m}$ is given in terms of birational transformations,
that is,
$T_j(x_{i,m}), T_j^{-1}(x_{i,m}) \in {\mathbb C}({\boldsymbol x})$
for any $i,j,m$.
\end{thm}

We shall illustrate this theorem with $I=\{1,2\}$ case.
Write
\[x_m:=x_{1,m}=t_1^{-1} T_2(v_{1,m}),  
\quad
y_m:=x_{2,m}=t_2^{-1} v_{2,m},  
\quad
x_m':=t_2^{-1} T_1(v_{2,m}), 
\quad 
y_m':=t_1^{-1} v_{1,m}.
\]
The compatibility condition of (\ref{eq:kp-lax})
yields ({\it cf.} (\ref{subeq:comp}))
\[
x_m y_m = x_m' y_m'
\quad \text{and} \quad
x_m+y_{m+1}=x_m'+y_{m+1}'
\quad
(m \in {\mathbb Z}/M {\mathbb Z})
\]
which provide a birational mapping
$r:(x,y) \to (x',y')$
of the form:
\[
x_{m}' = y_{m} \frac{P_{m-1}}{P_m}, \quad 
y_m'= x_m \frac{P_m}{ P_{m-1} },
\]
where $P_m$ is a polynomial in $(x,y)$ given as 
\[
P_m(x,y)= 
\sum_{a=1}^M 
\prod_{i=1}^{a-1} x_{m+i} \prod_{i=a+1}^M y_{m+i}.
\]
Note that in the above formula 
the suffixes $i$ of  
$x_i$ and $y_i$ are regarded 
as elements of ${\mathbb Z}/M {\mathbb Z}$,
namely, $x_{i+M}=x_i$ etc.
In addition, we introduce a permutation 
$\pi: x_m \leftrightarrow y_m$ of variables.
Apply $r \circ \pi$ to variables $x_m$ and $y_m$.
Therefore we can verify that
$r \circ \pi (x_m)=r(y_m)=y_m'=T_2^{-1}(x_m)=q T_1(x_m)$ and 
$r \circ \pi (y_m)=r(x_m)=x_m'=T_1(y_m)$.
Here notice that we have taken into account the similarity condition
$T_{1}T_{2}(v_{i,m})=v_{i,m}$ $(i=1,2)$.
Summarizing above,
the birational action of $T_1: {\mathbb C}({\boldsymbol x},{\boldsymbol y}) \circlearrowleft$ is given as follows:
\begin{equation} \label{eq:q-pa}
T_1(x_m)=q^{-1} x_m \frac{P_m}{ P_{m-1} }, \quad 
T_1(y_m)=y_m \frac{P_{m-1}}{ P_m } \quad 
(m \in {\mathbb Z}/M {\mathbb Z}).
\end{equation}
This discrete dynamical system looks $2M$-dimensional.
However,  
(i) if $M$ is odd, it possesses
$M+1$ conserved quantities 
 $x_my_m$ $(1 \leq m \leq M)$ and
$\prod_{i=1}^M x_i$;
(ii) if $M$ is even,
it does
$M+2$ ones
$x_my_m$,
$\prod_{i=1}^M x_i$
and 
$\prod_{j=1}^{M/2} x_{2j}/y_{2j-1}$.
Hence  the dimension of the dynamics (\ref{eq:q-pa}) is essentially $M-1$ (resp. $M-2$)
if $M$ is odd (resp. even).
As is known,
(\ref{eq:q-pa}) provides a $q$-analogue of the higher order Painlev\'e equation
of type  $A_{M-1}^{(1)}$ \cite{ny98} or the $M$-periodic closing of the Darboux chain \cite{adl, vs}.
In this paper we call (\ref{eq:q-pa})
the {\it $q$-Painlev\'e equation
of type  $A_{M-1}^{(1)}$}, denoted by $q$-$P(A_{M-1}^{(1)})$.
Note that  $q$-$P(A_{2}^{(1)})$ and $q$-$P(A_{3}^{(1)})$,
which are of two dimensions,
correspond to the fourth and fifth Painlev\'e equations, respectively.

\begin{remark}\rm
About Theorem~\ref{thm:kny}.
If $I=\{1,2, \ldots, L\}$, $J=\emptyset$ and the $M$-periodic condition is imposed,
the birational action of $T_i$ is generally governed by an affine Weyl group
of type $A_{L-1}^{(1)} \times A_{M-1}^{(1)}$;
see \cite{kny02a} and also \cite{kir}.
Note that the 
resulting discrete time flows $T_i$ $(1 \leq i \leq L)$ 
define a multi-variable version of a discrete Painlev\'e equation
which is called the {\it $q$-Painlev\'e system of type $(L,M)$}; 
see \cite[Sect.~3]{kny02b} for its explicit description.
It is still an interesting open question how to control, 
by the use of some Weyl groups, 
the time evolutions $T_i$
of $q$-KP 
(not to mention that of $q$-LUC)
in the case where $J \neq \emptyset$.
\end{remark}

\subsection{From $q$-KP/UC hierarchy to $q$-Painlev\'e equations: an overview}

As already mentioned,
a certain homogeneity and periodic constraint of $q$-KP
gives rise to 
a class of birational dynamical systems of
$q$-Painlev\'e type.
Analogously, it is known that $q$-LUC
also admits 
a similar type of reductions to some 
$q$-Painlev\'e equations.
Let us summarize the known results about
how $q$-KP or $q$-LUC corresponds to the $q$-Painlev\'e equations. 

Recall that $q$-KP possesses the following data:
$M$ and $(|I|,|J|)$, where
$M$ ($1 \leq M \leq \infty$)
is the order of periodicity on the dependent variables, {\it i.e.},
$\rho_{m+M}=\rho_m$
and $(|I|, |J|)$ specifies the set  
$\{ t_i \ (i \in I \cup J)\}$
of time variables.
The homogeneity constraint condition reads
\begin{equation}   \label{eq:homog-kp}
 \prod_{i \in I} T_i \prod_{j \in J} T_j^{-1} (\rho_m)=d_m \rho_m
\quad (d_m \in {\mathbb C}).
\end{equation}
As explained in Sect.~\ref{subsect:qkp}, 
if we choose 
$I=\{1,2\}$ and $J=\emptyset$ and $M$ ($\geq 3$: general),
then the time evolution of $T_1$ (or $T_2$) of $q$-KP 
yields the $q$-Painlev\'e equation of type $A_{M-1}^{(1)}$, 
denoted by $q$-$P(A_{M-1}^{(1)})$,
via the homogeneity constraint (\ref{eq:homog-kp}).
Also, it is known \cite{t06} that 
a $q$-analogue of the third Painlev\'e equation, $q$-$P_{\rm III}$, 
can be derived from $q$-KP
with $(|I|,|J|)=(3,0)$ and two-periodicity.
We sum up in Table 1 below the corresponding data of $q$-KP to each $q$-Painlev\'e equation:
\begin{center}
\small {\bf Table 1}. From $q$-KP to $q$-Painlev\'e equations
\\ 
\normalsize
\begin{tabular}{| c | c || l | c |}
\hline
$M$ : periodicity & $(|I|,|J|)$ & $T$ : time evolution of $q$-Painlev\'e equation
& Ref.
\\ 
\hline  
$M$ ($\geq 3$)  & $(2,0)$ & $T=T_1$ : $q$-$P(A_{M-1}^{(1)})$; 
$M=3 \Rightarrow $ $q$-$P_{\rm IV}$;
$M=4 \Rightarrow $ $q$-$P_{\rm V}$ & \cite{kny02b}
\\
\hline
2 & $(3,0)$ & $T=T_1$ : $q$-$P_{\rm III}$ & \cite{t06}
\\
\hline
\end{tabular}
\end{center}

Likewise, 
$q$-LUC has 
 the data:
$(M,N)$ and $(|I|,|J|)$, where
$M$ and $N$ ($1 \leq M, N \leq \infty$)
represent the period, {\it i.e.},
$\tau_{m+M,n}=\tau_{m,n+N}=\tau_{m,n}$
and $(|I|, |J|)$ specifies the set of time variables. 
Note that $q$-LUC with $N$ (or $M)=1$ is equivalent to  $q$-KP.  
The homogeneity constraint reads
\begin{equation}   \label{eq:homog-uc}
 \prod_{i \in I} T_i \prod_{j \in J} T_j^{-1} (\tau_{m,n})=d_{m,n} \tau_{m,n}
\quad (d_{m,n} \in {\mathbb C})
\end{equation}
with the balancing condition $d_{m,n}+d_{m+1,n+1}=d_{m,n+1}+d_{m+1,n}$; 
see Remark~\ref{rem:sym}.
The results for $q$-LUC are summarized as follows:
\begin{center}
\small {\bf Table 2}. From $q$-LUC to $q$-Painlev\'e equations
\\
\normalsize
\begin{tabular}{| c | c || l | c|}
\hline
$(M,N)$ : periodicity & $(|I|,|J|)$ &  $T$ : time evolution of $q$-Painlev\'e equation & Ref.
\\  \hline  
$(2,2)$  & $(2,2)$ &   $T=T_1T_2$ : $q$-$P_{\rm VI}$ & \cite{t05b, tm}
\\  \hline
$(3,3)$ & $(3,0)$ & $T=T_1T_2^{-1}$ : $q$-$P(E_6^{(1)})$  & \cite{t05b}
\\  \hline
$(M,M)$ $(M \geq 2)$ & $(2,2)$ & $T=T_1T_{-1}^{-1}$ (with $t_{-2}/t_{-1}=q^{1/2}$) : $q$-$P(A_{2M-1}^{(1)})$
& \cite{t05a}
\\ \hline
\end{tabular}
\end{center}
It is worth mentioning that
$q$-$P(A_{2M-1}^{(1)})$ $(M \geq 2)$ can be derived from $q$-KP or, 
alternatively, $q$-LUC.
We refer $q$-$P_{\rm VI}$ to be the $q$-analogue of the sixth Painlev\'e equation due to Jimbo--Sakai \cite{js}.
Note that $q$-$P_{\rm VI}$
can also be derived from the $q$-analogue of the three-wave resonant system, as shown by Kakei--Kikuchi \cite{kk}.

\begin{remark}\rm
We remember 
that the universal characters $S_{[\lambda,\mu]}$
(resp. the Schur functions $S_\lambda$) are homogeneous solutions of $q$-LUC
(resp. $q$-KP)
and readily compatible with the reduction constraints (\ref{eq:homog-uc}) (resp. (\ref{eq:homog-kp}))
under consideration;
see Prop.~\ref{prop:uc} and Remark~\ref{rem:sym}.
Hence it is immediate to construct special solutions of $q$-Painlev\'e equations in terms of 
$S_{[\lambda,\mu]}$ or $S_\lambda$;
see  \cite{kny02b, t05a, t05b, t06, tm} for details.
\end{remark}

In the rest of this paper, 
we will be devoted to
the reduction procedure from $q$-LUC
to certain birational dynamical systems of  $q$-Painlev\'e type.
Firstly, in Sect~\ref{subsect:j=0}, we deal with the case where 
$J=\emptyset$ and $(M,N)$ is general,
which in fact can be done in a similar manner as the $q$-KP case;
{\it cf}. Sect.~\ref{subsect:qkp}.
Secondly, in Sect~\ref{subsect:j=2},
we consider in particular the case where
$(|I|,|J|)=(2,2)$ and $(M,N)$ is general
and then present a higher-order analogue of  
$q$-$P_{\rm VI}$
as a result.

\subsection{The case of lattice $q$-UC hierarchy with $J=\emptyset$}
\label{subsect:j=0}

Assume $J=\emptyset$. 
In this case the argument can be proceeded along quite a 
parallel way
with the case of $q$-KP ({\it cf}. Sect.~\ref{subsect:qkp});
though we will here demonstrate only the case $|I|=2$ for simplicity.
Let $I=\{ 1, 2 \}$ and $J=\emptyset$.
Impose on the variables $\tau_{m,n}=\tau_{m,n}(t_1,t_2)$ 
the $(M,N)$-periodic condition:
$\tau_{m+M,n}=\tau_{m,n+N}=\tau_{m,n}$
and the homogeneity condition: 
$T_1T_2(\tau_{m,n})=q^{d_{m,n}} \tau_{m,n}$,
where $d_{m,n} \in {\mathbb C}$ are constant parameters
satisfying 
$d_{m,n}+d_{m+1,n+1}=d_{m,n+1}+d_{m+1,n}$.
Concerning the variables $u_{i,m,n}$,
the constraint above implies that
\begin{equation}  \label{eq:sim-u}
u_{i,m,n}=u_{i,m+M,n}=u_{i,m,n+N}
\quad
\text{and}
\quad
T_1T_2(u_{i,m,n})=q^{d_{m+1,n}+d_{m,n+2} -d_{m,n+1}-d_{m+1,n+1} } u_{i,m,n}.
\end{equation}
Introduce the dependent variables
\[
x_{m,n}:=t_1^{-1} T_2(u_{1,m,n}), \quad 
y_{m,n}:=t_2^{-1} u_{2,m,n+1},
\]
and also auxiliary variables
\[
x_{m,n}':=t_2^{-1} T_1(u_{2,m,n}), \quad 
y_{m,n}':=t_1^{-1} u_{1,m,n+1}.
\]
The compatibility condition of (\ref{eq:lax}) yields the formulae (recall (\ref{subeq:comp})):
\[
x_{m,n}y_{m,n}=x_{m,n}' y_{m,n}' \quad
\text{and} \quad
x_{m,n}+y_{m+1,n-1}=x_{m,n}'+y_{m+1,n-1}'
\quad (m \in {\mathbb Z}/M{\mathbb Z}, 
n \in {\mathbb Z}/N{\mathbb Z})
\]
which provide a birational mapping $r:(x,y) \mapsto (x',y')$ given by
\[
x_{m,n}'=y_{m,n} \frac{P_{m-1,n+1}}{P_{m,n}},
\quad
y_{m,n}'=x_{m,n} \frac{P_{m,n}}{P_{m-1,n+1}}.
\]
Here
\[
P_{m,n}(x,y)= \sum_{a=1}^{L} \prod_{i=1}^{a-1}
x_{m+i,n-i} \prod_{i=a+1}^L  y_{m+i,n-i}
\]
and $L$ is the least common multiple of $(M,N)$.
Prepare the mappings 
$\pi: x_{m,n} \leftrightarrow y_{m,n}$
and
$\sigma: (x_{m,n},y_{m,n}) \mapsto  (x_{m,n-1},y_{m,n+1})$.
Apply $r \circ \pi \circ \sigma$ to variables $x_{m,n}$ and $y_{m,n}$.
Therefore we see that
$r \circ \pi \circ \sigma ( x_{m,n})=r \circ \pi (x_{m,n-1})= r (y_{m,n-1}) = y_{m,n-1}' =T_2^{-1}(x_{m,n})$
and
$ r \circ \pi \circ \sigma ( y_{m,n})=r \circ \pi (y_{m,n+1})= r (x_{m,n+1}) = x_{m,n+1}' =T_1(y_{m,n})$.
In view of the similarity condition (\ref{eq:sim-u}), 
we observe that
$T_1(x_{m,n})=q^{d_{m+1,n}+d_{m,n+2} -d_{m,n+1}-d_{m+1,n+1} -1} T_2^{-1}(x_{m,n})$.
Finally, the birational action of 
$T_1: {\mathbb C}({\boldsymbol x},{\boldsymbol y}) \circlearrowleft$ 
turns out to be given as follows:
\begin{subequations}
\begin{align} 
T_1(x_{m,n})
&=q^{d_{m+1,n}+d_{m,n+2} -d_{m,n+1}-d_{m+1,n+1} -1}
 x_{m,n-1}  \frac{P_{m,n-1}}{P_{m-1,n}}, 
 \\
T_1(y_{m,n})
&=y_{m,n+1}\frac{P_{m-1,n+2}}{P_{m,n+1}}
\end{align}
\end{subequations}
for  $m \in {\mathbb Z}/M {\mathbb Z}$ and  $n \in {\mathbb Z}/N {\mathbb Z}$.

\subsection{The case of lattice $q$-UC hierarchy with $J \neq \emptyset$}
\label{subsect:j=2}

Let $I=\{ 1,2\}$ and $J=\{-1,-2\}$.
We impose on 
$\tau_{m,n}=\tau_{m,n}({\boldsymbol t})$ 
 the homogeneity condition:
$\tau_{m,n}(q {\boldsymbol t})=q^{d_{m,n}}  \tau_{m,n}({\boldsymbol t})$,
where $d_{m,n} \in {\mathbb C}$
fulfills the balance 
$d_{m,n}+d_{m+1,n+1}=d_{m,n+1}+d_{m+1,n}$.
Accordingly,
the variables $w_{m,n}=\tau_{m+1,n}/\tau_{m,n+1}$ satisfy
\begin{equation}
T_{1}T_{2} (w_{m,n})= c_{m,n} T_{-1}T_{-2}(w_{m,n})
\quad
\text{where}
\quad
c_{m,n}=q^{d_{m+1,n}-d_{m,n+1}}.
\end{equation}
We will often abbreviate the $q$-shift 
$T_{i_1} \cdots T_{i_r} T_{j_1}^{-1} \cdots T_{j_s}^{-1} (F)$
of a function $F=F({\boldsymbol t})$
to $F^{(i_1,\ldots,i_r) [j_1,\ldots,j_s]}$.
Let us now choose 
the dependent variables as
\begin{equation} \label{eq:def-fg}
f_{m,n}=\frac{ w_{m,n}^{(1)}}{ w_{m,n}^{(-1)}}
=\frac{\tau_{m+1,n}^{(1)} \tau_{m,n+1}^{(-1)}}{\tau_{m+1,n}^{(-1)} \tau_{m,n+1}^{(1)}   },
\quad 
g_{m,n}=\frac{ w_{m,n}^{(1,-1)}}{ w_{m,n}^{(-1,-2)}}
=\frac{\tau_{m+1,n}^{(1,-1)} \tau_{m,n+1}^{(-1,-2)}}{\tau_{m+1,n}^{(-1,-2)} \tau_{m,n+1}^{(1,-1)} }.
\end{equation}
Consider the field  
${\mathbb K}({\boldsymbol f}, {\boldsymbol g})$ of rational functions in $f_{m,n}$ and $g_{m,n}$ with
${\mathbb K}$ being a certain coefficient field. 

\begin{thm}   \label{thm:qp6}
The action of $T:=T_1T_2$ on variables $f_{m,n}$ and $g_{m,n}$ is given in terms of birational transformations,
that is,
$T^{\pm 1}(f_{m,n}), T^{\pm 1}(g_{m,n}) \in {\mathbb K}({\boldsymbol f}, {\boldsymbol g})$
for any $m,n$.
\end{thm} 
 
 \pf
 Recall (\ref{eq:w}) the functional equation satisfied by $w_{m,n}$:
\begin{equation}  \label{eq:ww}
w_{m,n}^{(i,j)} = \frac{w_{m,n+1}^{(i)}w_{m,n+1}^{(j)}}{ w_{m+1,n+1} }
\frac{t_i w_{m+1,n}^{(j)} - t_j w_{m+1,n}^{(i)} }{t_i w_{m,n+1}^{(j)} - t_j w_{m,n+1}^{(i)}}
\end{equation}
where  $i \neq j$ and  $i,j \in I \cup J =\{1,2,-1,-2\}$.

First we shall calculate $T_1T_2(f_{m,n})$.
 Apply $T_{-1}$ to (\ref{eq:ww}) with $(i,j)=(1,-2)$.
We then obtain
\begin{align} \nonumber
\frac{1}{c_{m,n}}T_{1}T_{2}(w_{m,n}^{(1)})
=
w_{m,n}^{(1,-1,-2)}  
&= \frac{w_{m,n+1}^{(1,-1)}w_{m,n+1}^{(-1,-2)}}{ w_{m+1,n+1}^{(-1)} }
\frac{t_{1} w_{m+1,n}^{(-1,-2)} - t_{-2} w_{m+1,n}^{(1,-1)} }{t_1 w_{m,n+1}^{(-1,-2)} - t_{-2} w_{m,n+1}^{(1,-1)}}
\\
&=
\frac{w_{m,n+1}^{(1,-1)} w_{m+1,n}^{(-1,-2)} }{ w_{m+1,n+1}^{(-1)}}
\frac{t_1- t_{-2}g_{m+1,n}}{t_1-t_{-2}g_{m,n+1}}.
\label{eq:fmn-1}
\end{align}
Applying $T_1$ to (\ref{eq:ww}) with $(i,j)=(2,-1)$, we have
\begin{align} \nonumber
T_1T_2(w_{m,n}^{(-1)})
=
w_{m,n}^{(1,2,-1)} 
&= \frac{w_{m,n+1}^{(1,2)}w_{m,n+1}^{(1,-1)}}{ w_{m+1,n+1}^{(1)} }
\frac{t_2 w_{m+1,n}^{(1,-1)} - t_{-1} w_{m+1,n}^{(1,2)} }{t_2 w_{m,n+1}^{(1,-1)} - t_{-1} w_{m,n+1}^{(1,2)}}
\\
\nonumber
&=
\frac{c_{m,n+1}w_{m,n+1}^{(-1,-2)} w_{m,n+1}^{(1,-1)}}{ w_{m+1,n+1}^{(1)} }
\frac{t_2 w_{m+1,n}^{(1,-1)} - t_{-1} c_{m+1,n}w_{m+1,n}^{(-1,-2)} }{t_2 w_{m,n+1}^{(1,-1)} - t_{-1}c_{m,n+1}w_{m,n+1}^{(-1,-2)}}
\\
&=
\frac{c_{m,n+1}w_{m+1,n}^{(-1,-2)} w_{m,n+1}^{(1,-1)}}{ w_{m+1,n+1}^{(1)} }
\frac{t_2 g_{m+1,n}-t_{-1}c_{m+1,n} }{ t_2  g_{m,n+1} -t_{-1} c_{m,n+1} }.
\label{eq:fmn-2}
\end{align}
Note that we have used $w_{m,n}^{(1,2)}=c_{m,n}w_{m,n}^{(-1,-2)}$
between the first and second lines.
Combining (\ref{eq:fmn-1}) with (\ref{eq:fmn-2}) leads to
\begin{equation}  \label{eq:fmn}
\frac{T_1T_2(f_{m,n})}{f_{m+1,n+1}}
= 
\frac{c_{m,n}}{ c_{m,n+1}}  
\frac{ \left(g_{m+1,n}- \frac{t_1}{t_{-2}}\right)  \left(g_{m,n+1} -c_{m,n+1} \frac{t_{-1}}{t_2} \right) }{ \left(g_{m,n+1}- \frac{t_1}{t_{-2}} \right) \left( g_{m+1,n}- c_{m+1,n}\frac{t_{-1}}{t_2} \right)}.
\end{equation}

Next we shall concern $(T_1T_2)^{-1}(g_{m,n})$.
Notice that $w_{m+1,n+1}=c_{m+1,n+1} (T_1T_2)^{-1} (w_{m+1,n+1}^{(-1,-2)})$.
It  therefore  follows from (\ref{eq:ww})
with $(i,j)=(1,-1)$ that
\begin{equation} \label{eq:gmn-1}
(T_1T_2)^{-1} (w_{m+1,n+1}^{(-1,-2)})= \frac{1}{c_{m+1,n+1}} 
\frac{w_{m,n+1}^{(1)}w_{m+1,n}^{(-1)}}{ w_{m,n}^{(1,-1)} }
\frac{t_1-t_{-1}f_{m+1,n}}{t_1-t_{-1}f_{m,n+1}}.
\end{equation}
By applying $T_2^{-1}T_{-1}$
to  (\ref{eq:ww}) with $(i,j)=(2,-2)$,
we have
\begin{equation} \label{eq:(2,-2)}
w_{m,n}^{(-1,-2)} = \frac{w_{m,n+1}^{(-1)}w_{m,n+1}^{(-1,-2) [2]}}{ w_{m+1,n+1}^{(-1)[2]} }
\frac{q^{-1}t_2 w_{m+1,n}^{(-1,-2)[2]} - t_{-2} w_{m+1,n}^{(-1)} }{q^{-1}t_2 w_{m,n+1}^{(-1,-2)[2]} - t_{-2} w_{m,n+1}^{(-1)}}.
\end{equation}
Observe that
$w_{m,n}^{(-1,-2)[2]}=w_{m,n}^{(1)}/c_{m,n}$
and
$w_{m+1,n+1}^{(-1)[2]}=
(T_1T_2)^{-1}
w_{m+1,n+1}^{(1,-1)}$.
We then verify from (\ref{eq:(2,-2)}) that  
\begin{equation}  \label{eq:gmn-2}
(T_1T_2)^{-1}
w_{m+1,n+1}^{(1,-1)}
=
\frac{1}{c_{m+1,n}}
\frac{w_{m+1,n}^{(-1)} w_{m,n+1}^{(1)}}{ w_{m,n}^{(-1,-2)} }
\frac{t_2 f_{m+1,n} - q c_{m+1,n} t_{-2}}{t_2f_{m,n+1}-q c_{m,n+1} t_{-2}}.
\end{equation}
If we put (\ref{eq:gmn-1}) and (\ref{eq:gmn-2}) together,
we arrive at 
\begin{equation} \label{eq:gmn}
\frac{g_{m,n}}{(T_1T_2)^{-1}(g_{m+1,n+1})}
=
\frac{c_{m+1,n}}{c_{m+1,n+1}}
\frac{ \left(f_{m+1,n}-  \frac{t_1}{t_{-1}} \right)\left(f_{m,n+1}-q c_{m,n+1} \frac{t_{-2}}{t_2} \right) }{ \left(f_{m,n+1} -\frac{t_1}{t_{-1}} \right) \left(f_{m+1,n} - q c_{m+1,n} \frac{t_{-2}}{t_2} \right)}.
\end{equation}

Finally, by virtue of (\ref{eq:fmn}) and (\ref{eq:gmn}),
it is clear that $T=T_1T_2$ acts on $f_{m,n}$ and $g_{m,n}$ 
as a birational mapping.
\qed
\\

Let us slightly refine the birational dynamics,
(\ref{eq:fmn}) and (\ref{eq:gmn}), constructed above in Theorem~\ref{thm:qp6}.
Put
\[
\alpha= \frac{t_1}{t_{-2}}, \quad \beta= \frac{t_{-1}}{t_2}, \quad
\gamma=\frac{t_1}{t_{-1}}, \quad \delta= \frac{t_{-2}}{t_2}.
\]
We have then
a birational dynamical system 
$T:(\alpha,\beta,\gamma,\delta;f_{m,n},g_{m,n}) \mapsto
\left(q \alpha,\beta/q,q \gamma,\delta/q; \overline{f_{m,n}}, \overline{g_{m,n}} \right)$,
\begin{subequations} \label{subeq:qp6}
\begin{align}
\overline{f_{m,n}}
&=
\frac{c_{m,n}}{c_{m,n+1}}  
\frac{ (g_{m+1,n}-\alpha)  (g_{m,n+1} -c_{m,n+1} \beta) }{(g_{m,n+1}- \alpha)( g_{m+1,n}- c_{m+1,n}\beta)}f_{m+1,n+1},
\\
\overline{g_{m,n}}
&=
\frac{c_{m+1,n}}{c_{m+1,n+1}}
\frac{ \left(\overline{f_{m+1,n}}- q \gamma \right) \left(\overline{f_{m,n+1}}- c_{m,n+1}\delta \right) }{ \left(\overline{f_{m,n+1}} - q \gamma \right) \left(\overline{f_{m+1,n}} -  c_{m+1,n} \delta\right)}
g_{m+1,n+1},
\end{align}
\end{subequations}
with
$\alpha \delta/\beta \gamma=1$
and
$c_{m,n}=q^{d_{m+1,n}-d_{m,n+1}}$.

From now on,
we shall impose the $(M,N)$-periodicity on the suffixes $(m,n)$ of the variables. 
Let $L$ denote the least common multiple of $(M,N)$,
and
recall (\ref{eq:def-fg}).
Hence the dynamics (\ref{subeq:qp6}) turns out to possess
the $2MN/L$ conserved quantities:
\begin{equation}
\prod_{i=1}^{L} f_{m+i,n-i}=\prod_{i=1}^{L} g_{m+i,n-i}= 1.
\end{equation}
Moreover, if $M=N=2$ then the dynamics (\ref{subeq:qp6}) is actually
closed in two variables, {\it e.g.}, 
$f=f_{1,1}$ and $g=g_{1,2}$:
\begin{align*}
\overline{f}  
&=
\frac{c_{1,1}}{f}  
\frac{ (\alpha g-1)  (g -c_{1,2} \beta) }{(g- \alpha)( \beta g - c_{1,2})},
\\
\overline{g} 
&=
\frac{c_{1,2}}{g}
\frac{ \left( q \gamma \overline{ f } -1 \right) \left(\overline{f}- c_{1,1}\delta \right) }{ \left(\overline{f} - q \gamma \right) \left(\delta \overline{ f } -c_{1,1} \right)},
\end{align*}
where
$(\overline{\alpha},\overline{\beta},\overline{ \gamma },\overline{\delta})
=
(q \alpha,\beta/q,q \gamma,\delta/q)$
and
$\alpha \delta/\beta \gamma=1$.
This coincides with 
the $q$-analogue of the sixth Painlev\'e equation ($q$-$P_{\rm VI}$);
{\it cf.} \cite{js}.
For this reason
we regard  
(\ref{subeq:qp6}) 
as a higher-order extension of $q$-$P_{\rm VI}$.

\small
\paragraph{\it Acknowledgement.}
I would like to thank 
Saburo Kakei, 
Tetsuya Kikuchi, 
Frank Nijhoff, 
Masatoshi Noumi 
and Yasuhiko Yamada for valuable discussions. 
This work was partly conducted during my stay in the 
Issac Newton Institute of Mathematical Sciences 
for 
program
"Painlev\'e Equations and Monodromy Problems"
(2006).
My research is supported by
JSPS Grant 19840039
and the grant for Basic Science Research Projects of the Sumitomo Foundation
071254.


\begin{thebibliography}{99}

\bibitem{adl}
Adler, V.E.:
Nonlinear chains and Painlev\'e equations.
Phys. D {\bf 73}, 
335--351
(1994)


\bibitem{js}
Jimbo, M., Sakai, H.:
A $q$-analog of the sixth Painlev\'e equation.
Lett. Math. Phys. {\bf 38},
145--154  (1996)



\bibitem{kny02a}
Kajiwara, K.,  Noumi, M.,  Yamada, Y.: 
Discrete dynamical systems with $W(A_{m-1}^{(1)} \times A_{n-1}^{(1)})$ symmetry.
Lett. Math. Phys. {\bf 60},
211--219
(2002)

 
\bibitem{kny02b}
Kajiwara, K., Noumi, M.,  Yamada, Y.:
$q$-Painlev\'e systems arising from $q$-KP hierarchy.
Lett. Math. Phys. {\bf 62},  
259--268 (2002)


\bibitem{kk}
Kakei, S., Kikuchi, T.:
A $q$-analogue of 
$\widehat{\mathfrak{g l}}_ 3$ 
hierarchy and $q$-Painlev\'e VI.  
J. Phys. A  {\bf 39},   
12179--12190 (2006) 

\bibitem{kir}
Kirillov, A.N.:
Introduction to tropical combinatorics.
In: 
{\it
Physics and combinatorics, 2000},
Kirillov, A.N.,  Liskova, N. (eds.),
World Scientific,  Singapore,
2001,
pp. 82--150

\bibitem{koi} 
Koike, K.:
On the decomposition of tensor products of 
the representations of the classical groups:
By means of the universal characters.
Adv. Math. {\bf 74}, 
57--86 (1989)

\bibitem{ny98}
Noumi, M., Yamada, Y.:
Higher order Painlev\'e equations of type $A_l^{(1)}$.
Funkcial. Ekvac. {\bf 41},
483--503
(1998)


\bibitem{sat}
Sato, M.:
Soliton equations as dynamical systems on an infinite dimensional Grassmann manifold.
RIMS Koukyuroku {\bf 439},
30--46  (1981)



\bibitem{t04}
Tsuda, T:
Universal characters and an extension of the KP hierarchy.
Comm. Math. Phys. {\bf 248}, 
501--526 (2004)


\bibitem{t05a}
Tsuda, T.:
Universal characters and $q$-Painlev\'e systems.
Comm. Math. Phys.  {\bf 260},  
59--73 (2005)



\bibitem{t05b}
Tsuda, T.:
Universal character and $q$-difference Painlev\'e equations 
with affine Weyl groups.
(Preprint,  UTMS 2005--21;
available also from arXiv:0811.3112)

\bibitem{t06}
Tsuda, T.:
Tau functions of $q$-Painlev\'e III and IV equations.
Lett. Math. Phys. {\bf 75},
39--47 (2006)


\bibitem{tm}
Tsuda, T.,  Masuda, T.:
$q$-Painlev\'e VI equation arising from $q$-UC hierarchy.
Comm. Math. Phys. {\bf 262}  
595--609 (2006)


\bibitem{vs}
Veselov, A.P.,  Shabat, A.B.:
A dressing chain and the spectral theory of the Schr\"odinger operator.
Funct. Anal. Appl. {\bf 27}, 
81--96 (1993)

\end{thebibliography}
\end{document}